\documentclass[twocolumn,showpacs,preprintnumbers,amsmath,amssymb,pra]{revtex4}

\usepackage{array}
\usepackage{graphicx} 
\usepackage{dcolumn}
\usepackage{bm}
\usepackage{amsmath}
\usepackage{enumitem}

\newcommand{\opr}[1]{\ensuremath{\hat{#1}}}

\newcommand{\abs}[1]{\ensuremath{\left|#1\right|}}

\newcommand{\sign}[1]{\ensuremath{\mbox{sgn}\left(#1\right)}}

\newcommand{\ket}[1]{\ensuremath{\left|\left. #1\right.\right>}}
\newcommand{\bra}[1]{\ensuremath{\left<\left.#1\right.\right|}}
\newcommand{\braket}[2]{\ensuremath{\left<\left.#1\right|#2\right>}}

\newcommand{\expecv}[2]{\ensuremath{\left<#1\left|#2\right|#1\right>}}

\begin{document}


\title{Synchronizing quantum and classical clocks made of quantum particles}
\author{Philip Caesar M. Flores}
\email{pflores@nip.upd.edu.ph}

\author{Roland Cristopher F. Caballar}
\email{rfcaballar@up.edu.ph}

\author{Eric A. Galapon}
\email{eric.galapon@up.edu.ph}

\affiliation{Theoretical Physics Group, National Institute of Physics, University of the Philippines, Diliman Quezon City, 1101 Philippines}


\begin{abstract}

We demonstrate that the quantum corrections to the classical arrival time for a quantum object in a potential free region of space, as computed in Phys. Rev. A {\bf 80}, 030102(R) (2009), can be eliminated up to a given order of $\hbar$ by choosing an appropriate position-dependent phase for the object's wave-function This then implies that we can make the quantum arrival time of the object as close as possible to its corresponding classical arrival time, allowing us to synchronize a classical and quantum clock which tells time using the classical and quantum arrival time of the object, respectively . We provide an example for synchronizing such a clock by making use of a quantum object with a position-dependent phase imprinted on the object's initial wave-function with the use of an impulsive potential.
\end{abstract}
%
%
\maketitle

\section{Introduction}

Time's status as a quantum mechanical observable has been a subject of intense study in recent years \cite{pauli,mugarev,mugabook,mugabook2,halliwell,yearsley,winful,gal,gal3,denny,erictunn,bunaoeric,caballarPLA,galaponWS,galaponSpringer,bunaoeric2}. This attention given to the role of time in quantum mechanics has allowed the development of models for quantum clocks which are described mathematically by quantum time distributions or equivalently by the expectation values of the time operators corresponding to these distributions. Muga et al \cite{muga,muga3,hegerfeldt} have shown one way by which this can be done using a two-level atom moving in a potential-free region incident on a transverse optical field at a point $q=q_o$. In this scheme, the instant of time when the atom emits a photon due to its excitation and relaxation as a result of its interaction with the optical field is the arrival time of the atom at the optical field's location. The resulting arrival time distribution for this system was shown to be identical to the time of arrival distribution obtained by Kijowski \cite{kijowski} from the non-self-adjoint free particle arrival time operator formulated by Bohm and Aharonov \cite{aharonov}, as well as to the distribution corresponding to the self-adjoint arrival time operator for a confined particle in the limit as the length of the confining region approaches infinity \cite{galmuga}.

However, it is possible to use other types of systems to construct a quantum clock by measuring its arrival time at a given point. As an example, we can use neutrons moving in free space, with neutrons having the advantage of having been used in numerous experiments as a probe to test the validity of quantum mechanics, as well as to investigate the effects of gravity on quantum systems \cite{rauch}. One way by which a quantum clock can be constructed using neutrons is to make use of a time-of-flight spectroscopy setup, wherein the time of flight of neutrons scattered off of a target is measured \cite{rauch,brockhouse}. The resulting time of flight of these neutrons is identical to the time of arrival of these neutrons at a given point $q_o$ in space where the time of flight is measured. Compared to a quantum clock constructed using Muga et al's two-level atom setup, this type of quantum clock eliminates the need for additional equipment to detect the arrival of the particle; all one needs is a particle detector positioned around the neutron source to register the arrival of the particle, and the instant the detector clicks is the neutron's arrival time at that point. Classically, this time of arrival of the scattered neutrons is given as $t=-q_o/v_o$, where we have chosen a coordinate system such that the arrival point is at the origin, $q_o$ is the position of the neutron relative to the scattering target, and $v_o$ is the velocity. 

In time-of-flight experiments involving quantum objects, it is assumed that the time-of-flight is to be computed classically and at the same time each object is a quantum wave-packet. Obviously these two assumptions are inconsistent because the former assumes each object to be localized while the latter assumes each object to be delocalized in space. However, it is the consensus that these two assumptions are consistent thus, in Ref. \cite{gal2}, Galapon examined the consistency of the two assumptions. In particular, Galapon treated the time of arrival of scattered neutrons in a time of flight experiment quantum mechanically using the general theory of time-of-arrival operator in the presence of an interaction potential formulated in Ref. \cite{gal4}, assuming that the initial quantum state of the neutron is a quantum wave-packet and that the neutron is moving through a potential-free region of space. This quantum mechanical treatment of the neutron's time of arrival shows that the size of the initial wave-packet introduces measurable quantum corrections to the expected classical time of arrival, and that consistency can only be maintained under certain conditions. Moreover, Galapon has shown that the usual assumptions used in time of flight experiments is only consistent with wave-packets of constant phase with a leading quantum correction term in the order of $O(\hbar^2)$ while for wave-packets with a position dependent phase, the leading quantum correction term is in the order of $O(\hbar)$. 

The resulting quantum corrections depend on a variety of parameters corresponding to the wave-packet, such as the initial position and momentum as well as the size and mass of the wave-packet. These quantum corrections imply that a quantum clock constructed using neutrons and whose time interval marker is its quantum arrival time may run slower or faster than a classical clock which makes use of neutrons as well, but which uses the neutrons' classical arrival time as its time interval marker. Generalizing, this result implies that if one constructs a quantum clock using particles, and makes use of their corresponding arrival times as time interval markers for this clock, then a classical clock and a quantum clock will not run at the same rate. However, it is still possible to synchronize these two clocks by minimizing the effects of the quantum correction terms to the classical time of arrival by introducing a position dependent phase that will make the quantum correction terms vanish. To do so, we show that the resulting quantum correction terms are functions of the position-dependent phase and introduce certain conditions to the phase that will make these quantum correction terms vanish up to an order $n$ of $\hbar$. This means that by introducing a position-dependent phase onto the object's initial wave-function we are eliminating the dependence of the expectation value of a particle's quantum arrival time on all parameters specific to its corresponding quantum wave-packet, except for the expectation value of its initial position and momentum. This simplifies the quantum arrival time's expectation value and brings it in line with the quantum particle's corresponding classical arrival time, which depends only on its initial position and momentum. Thus, we are able to synchronize quantum and classical clocks in the sense that we are eliminating the quantum correction terms in the order of $O(\hbar^3)$ or higher in order to increase the accuracy of the calculated time of flight in TOF experiments.

The rest of the paper is structured as follows. Section \ref{timeoperator} provides details on how the quantum arrival time of a scattered object moving in free space is computed, together with the corresponding correction terms. Section \ref{vanishing} shows how these quantum correction terms can be eliminated by introducing a position-dependent phase to the quantum wave-function of the particle. Section \ref{implementation} demonstrates how the quantum clock can be physically implemented. Section \ref{example} describes a specific example. Section \ref{summary} summarizes the paper and outlines the conclusions.

\section{The use of arrival time operators to describe a quantum clock}\label{timeoperator}

Consider a beam of classical non-interacting free identical particles each with mass $\mu$ and kinetic energy $E_o$.  Let the distribution of the initial positions of the particles be given by $\Phi(q)$ at $t=0$ so that the classical expected time of arrival, say at the origin, for such a system is given by 
\begin{equation}
\bar{\tau}_{class}=-\int_{-\infty}^{+\infty}\frac{q}{v_o} \Phi(q) dq=-\frac{q_o}{v_o},
\label{tclass}
\end{equation}
where $q_o=\int_{-\infty}^{+\infty}q \Phi(q) dq$ is the mean initial position of the particles and $v_o=\sqrt{2 E_o/\mu}$ is the velocity of each particle \cite{gal2}.

Now consider a beam of quantum non-interacting identical particles, each with mass $\mu$ and described by one particle wave-functions $\{\phi(q)\}$ in the system Hilbert space. Its quantum arrival time is described in terms of the expectation value of the quantum arrival time operator for a free particle, which has the explicit form
\begin{align}
(\mathbf{T}\phi)(q)=&\frac{\mu}{4i\hbar}\int_{-\infty}^{\infty}(q+q')\mathrm{sgn}(q-q')\phi(q')dq'\nonumber\\
=&\frac{\mu}{4i\hbar}\int_{-\infty}^{q}(q+q')\phi(q')dq'\nonumber\\
&-\frac{\mu}{4i\hbar}\int_{q}^{\infty}(q+q')\phi(q')dq'.
\label{freetoa}
\end{align}
The expectation value of the quantum arrival time will then have the explicit form
\begin{align}
\bar{\tau}_{quant}=&\bra{\phi}\opr{T}\ket{\phi}\nonumber=\int_{-\infty}^{\infty}\bar{\phi}(q)(\mathbf{T}\phi)(q)dq\nonumber\\
=&\frac{\mu}{4i\hbar}\int_{-\infty}^{\infty}\int_{-\infty}^{\infty}\bar{\phi}(q)(q+q')\mathrm{sgn}(q-q')\phi(q')dq'dq\nonumber\\
=&\frac{\mu}{4i\hbar}\int_{-\infty}^{\infty}\int_{-\infty}^{q}(q+q')\bar{\phi}(q)\phi(q')dq'dq\nonumber\\
&-\frac{\mu}{4i\hbar}\int_{-\infty}^{\infty}\int_{q}^{\infty}(q+q')\bar{\phi}(q)\phi(q')dq'dq.
\label{freetoaexp}
\end{align}

We note that the one particle wave-functions $\{\phi(q)\}$ in the system Hilbert space must satisfy the following properties: (i) the expectation value of the time of arrival operator reduces to the classical value in the limit as $\hbar$ approaches zero, and (ii) the quantum momentum expectation value is equal to the classical momentum of the incident particle $\expecv{\phi}{\opr{p}}=\mu v_o$ where $\opr{p}$ is the momentum operator. The second condition guarantees that the wave-packet moves abreast with its classical counterpart \cite{gal2}. 

Given that the expectation value of the time of arrival operator for the beam of particles exists, the wave-functions of the class assume the form $\phi(q)=\varphi(q) e^{i k q}$, where $k=\mu v_o/\hbar$,  and $\varphi(q)$ is a wave-packet that is (i) independent of $\hbar$, (ii) satisfies $|\varphi(q)|^2=\Phi(q)$, and (iii) has zero momentum expectation value. The third condition implies that
\begin{equation}
\int_{-\infty}^{+\infty}\bar{\varphi}(q) \frac{d \varphi (q)}{dq} dq=0.
\label{propphi3}
\end{equation}
From Ref. \cite{gal2} the wave-packet, $\varphi(q)$, has the form $\varphi(q)=e^{i\vartheta(q)}\sqrt{\Phi(q)}$ such that it satisfies the three properties stated above. Substituting $\varphi(q)=e^{i\vartheta(q)}\sqrt{\Phi(q)}$ to Eq. \eqref{propphi3} yields the equality
\begin{equation}
\int_{-\infty}^{+\infty}\Phi(q)\frac{d\vartheta(q)}{dq}dq=0,
\label{cond1}
\end{equation}
where $\vartheta(q)$ is generally a real-valued function of $q$. The equality in Eq. \eqref{cond1} follows from the assumed differentiability of $\Phi(q)$ and the finiteness of $\bar{\tau}_{class}$. If $f(q)$ is a function orthogonal to $\Phi(q)$, then the phase is given by $\vartheta(q)=\int f(q) dq$. There are infinitely many functions $f(q)$ orthogonal to $\Phi(q)$ because the finiteness of $\bar{\tau}_{class}$ implies that $\Phi(q)$ belongs to the system Hilbert space, which is infinite dimensional. Thus, there are infinitely many functions orthogonal to $\Phi(q)$ because every vector in the Hilbert space has infinitely many orthogonal vectors.

The quantum image of the classical beam of particles now comprises all wave-functions of the form
\begin{equation}
\phi(q)=e^{i \vartheta(q) + i \sqrt{2\mu E_0} q/\hbar}\sqrt{\Phi(q)},
\label{wave1}
\end{equation}
where $\vartheta(q)$ is an arbitrary real valued function with $\vartheta'(q)$ orthogonal to $\Phi(q)$.  From \cite{gal2} the full asymptotic form of the expectation value of the quantum time of arrival (QTOA) $\bra{\phi}\opr{T}\ket{\phi}$ in terms of the kinetic energy $E_o$ is
\begin{eqnarray}\label{asymptotic}
\bar{\tau}_{quant}
\sim  - \sqrt{\frac{\mu}{2 E_o}}\!\! \sum_{n=0}^{\infty} \!\!\frac{\hbar^{2n} \chi_1^{(n)}}{(2\mu E_o)^n} 
\!+\! \frac{1}{2 E_o}\!\! \sum_{n=0}^{\infty}  \!\! \frac{(-1)^n\hbar^{2n+1} \chi_2^{(n)}}{(2\mu E_o)^{n}}, \label{asymp}
\end{eqnarray}
as $(\hbar^2/\mu E_o)\rightarrow 0$, where
\begin{eqnarray}
\chi_1^{(n)}&=&\int_{-\infty}^{+\infty}q |\varphi^{(n)}\!(q)|^2 dq,
\label{chi1}
\end{eqnarray}
and
\begin{eqnarray}
\chi_2^{(n)}&=&\int_{-\infty}^{+\infty} q \, \mbox{Im}\!\left[\bar{\varphi}(q) \varphi^{(2n+1)}\!(q)\right] dq.
\label{chi2}
\end{eqnarray}
Generally, the right hand side of Eq. \eqref{asymp} diverges but meaningful numerical values can be extracted through super-asymptotic summation \cite{bender, balser}. These finite numerical values from Eq. \eqref{asymp} then imply that the quantum arrival time for the particle may be equal, less than, or greater than the classical arrival time depending on the form of the incident wave-function. 

We note that the dispersion of the wave-packet corresponding to a quantum particle as it evolves over time is the physical mechanism that gives rise to the nonzero correction terms in the expectation value of its arrival time. To illustrate this, consider a quantum particle moving along a line towards a particle detector located at some point $x$. At a given instant of time $t$, the particle will have a finite probability of being found in the neighborhood of $x$ (where the detector is located). However, it will also have a finite probability of being found at a point $x'$ elsewhere along the line, thus reducing the probability that the particle is at point $x$, and increasing the probability that it will be found elsewhere along the line. This implies that the expectation value for the position of the particle will change, and so too will the expectation value of the momentum, and so too as a result will the expectation value of the quantum arrival time of the particle at point $x$. 

The question now arises as to how we can mitigate the effects of the quantum corrections to the expected QTOA in order to synchronize our classical and quantum clocks. In the following section, we outline and demonstrate how this can be achieved up to some order of $\hbar$ by an appropriate choice of the phase $\vartheta(q)$ in Eq. \eqref{wave1}. 

\section{Vanishing of the quantum correction terms}\label{vanishing}

Using the exact expression for the expectation value of the QTOA given in Eq. \eqref{freetoaexp}, it is not apparent how the wave-function given by Eq. \eqref{wave1} can be modified in such a way that the expectation value of the QTOA is as close as we wish to the classical value. However, the asymptotic expansion of the expected QTOA  given by Eq. \eqref{asymptotic} provides a means of extracting conditions on the wave-function $\phi(q)$ to bring the expectation value of the QTOA nearer to the classical value to some order of $\hbar$. The basic idea is to choose the phase $\vartheta(q)$ for a given $\Phi(q)$ such that a predetermined number, $N$, of the leading quantum correction terms vanish so that $\bar{\tau}_{quant}=\bar{\tau}_{class}+\mathcal{O}(\hbar^{N+1})$. This requires $\chi_1^{(l)}=0$ for $l=1,\dots,\lfloor N/2\rfloor$, and $\chi_2^{(m)}=0$ for $m=0, \dots, \lfloor N/2\rfloor +1$, in addition to the condition \ref{cond1}. In this paper, we solve for the cases $N=1$ and $N=2$, so that $\bar{\tau}_{quant}=\bar{\tau}_{class}+\mathcal{O}(\hbar^2)$ and $\bar{\tau}_{quant}=\bar{\tau}_{class}+\mathcal{O}(\hbar^{3})$, respectively.

In the rest of the paper, we will consider Gaussian wave-functions described by
\begin{equation}
\psi_{np}(q)=\frac{1}{(\sigma \sqrt{2 \pi})^{1/2}}e^{-\frac{(q-q_0)^2}{4\sigma^2}}e^{i \sqrt{2\mu E_0}q/\hbar},
\label{nophase}
\end{equation}
\begin{equation}
\psi_{wp}(q)=\frac{1}{(\sigma \sqrt{2 \pi})^{1/2}}e^{-\frac{(q-q_0)^2}{4\sigma^2}}e^{i \sqrt{2\mu E_0}q/\hbar}e^{i \vartheta(q)}
\label{withphase},
\end{equation}
where the subscripts $\text{\lq \lq np \rq \rq}$ indicates that the wave-function has no position dependent phase and $\text{\lq \lq wp \rq \rq}$ indicates otherwise. Both of these wave-functions correspond to the same probability density
\begin{equation}\label{density}
\Phi(q)=\frac{1}{\sigma \sqrt{2\pi}} e^{-(q-q_0)^2/2\sigma^2}.
\end{equation}
From \cite{gal2}, the wave-function $\psi_{np}(q)$ has a leading order correction term $\mathcal{O}(\hbar^2)$ to the classical time of flight; and the wave-function $\psi_{wp}(q)$ has a leading $\mathcal{O}(\hbar)$ correction term. 

Now if we want to improve on the order $\mathcal{O}(\hbar^2)$ of $\psi_{np}(q)$, we have to modify it by introducing a phase, leading to the wave-function $\psi_{wp}(q)$, which will in turn lead to an order $\mathcal{O}(\hbar)$ correction. Let us tackle eliminating the $\mathcal{O}(\hbar)$ correction so that $\bar{\tau}_{quant}=\bar{\tau}_{class}+\mathcal{O}(\hbar^2)$. Later we will use our solution here to eliminate the $\mathcal{O}(\hbar^2)$ correction. There are two conditions to satisfy, Eq. \eqref{cond1} and $\chi_2^{(0)}=0$. The second condition translates to
\begin{equation}\label{cond2}
\int_{-\infty}^{\infty} q \Phi(q) \frac{\mathrm{d}\vartheta}{\mathrm{d}q}\, \mathrm{d}q=0 ,
\end{equation}
which ensures the vanishing of the $\mathcal{O}(\hbar)$ correction term. Notice that Eqs. \eqref{cond1} and \eqref{cond2} are both linear in the phase $\vartheta(q)$. This observation will be important later in our elimination of the $\mathcal{O}(\hbar^2)$ correction term.

As there are two (linear) conditions to satisfy, we assume that the phase is of the form
\begin{equation}
\vartheta(q)=c_1 \vartheta_{1}(q)+c_2 \vartheta_{2}(q),
\label{sol1}
\end{equation}
where $c_1$ and $c_2$ are real valued constants to be determined and $\vartheta_1(q)$ and $\vartheta_2(q)$ are linearly independent real valued functions. Substituting Eq. \eqref{sol1} back into Eqs. \eqref{cond1} and \eqref{cond2}, we obtain the following linear system of $2$ equations with $2$ unknowns, 
\begin{align} \nonumber
c_1\int_{-\infty}^{+\infty}\Phi(q)\dfrac{d\vartheta_1(q)}{dq}dq+c_2\int_{-\infty}^{+\infty}\Phi(q)\dfrac{d\vartheta_2(q)}{dq}dq= & 0 \\
c_1\int_{-\infty}^{+\infty}q \Phi(q)\dfrac{d\vartheta_1(q)}{dq}dq+c_2\int_{-\infty}^{+\infty}q \Phi(q)\dfrac{d\vartheta_2(q)}{dq}dq=& 0.
\label{linear}
\end{align}
Provided the determinant of the coefficients vanish, solutions for  $c_1$ and $c_2$ exist.

Incidentally the system of equations, Eq. \eqref{linear}, is exactly the same set of equations addressed and solved in \cite{bunao}. For the given Gaussian probability density, Eq. \eqref{density}, phases with definite parity can be obtained. They are given by
\begin{align} \nonumber
\label{odd}
\vartheta_{odd}(q)=&\dfrac{2\sigma\sqrt{\pi}}{2^l 2^m l!}\sqrt{\dfrac{(2l)!}{(2m)!}}\dfrac{\sigma}{4m+2}H_{2m+1}(\frac{q-q_{0}}{\sigma})\\
&-\dfrac{2\sigma\sqrt{\pi}}{2^l 2^m m!}\sqrt{\dfrac{(2m)!}{(2l)!}}\dfrac{\sigma}{4l+2}H_{2l+1}(\frac{q-q_{0}}{\sigma}) \\
\label{even} \nonumber
\vartheta_{even}(q)=&\dfrac{2\sigma\sqrt{\pi}}{2^l 2^m l!}\sqrt{\dfrac{(2l+1)!}{(2m+1)!}}\dfrac{\sigma}{4m+4}H_{2m+2}(\frac{q-q_{0}}{\sigma})\\
&-\dfrac{2\sigma\sqrt{\pi}}{2^l 2^m m!}\sqrt{\dfrac{(2m+1)!}{(2l+1)!}}\dfrac{\sigma}{4l+4}H_{2l+2}(\frac{q-q_{0}}{\sigma})
\end{align}
where $H_j(x)$ is the Hermite polynomial and $l\neq m$. Any of these phases will eliminate the $\mathcal{O}(\hbar)$ correction term but not necessarily the higher order correction terms.

\begin{figure*}
	\includegraphics[width=0.8\textwidth]{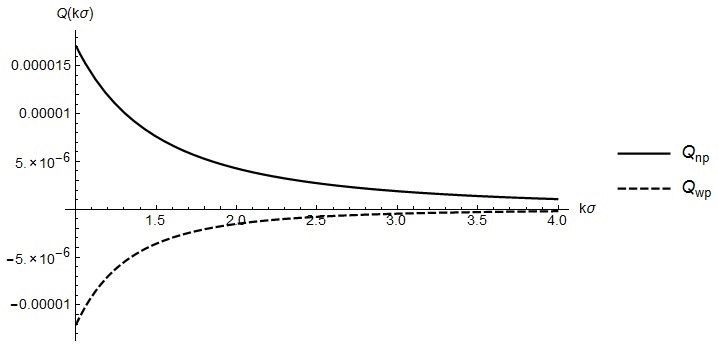}
	\caption{Comparison of the leading quantum correction term factor $Q_{np}$ and $Q_{wp}$ for $\psi_{np}$ and $\psi_{qp}$, respectively, with parameters $\alpha=5(4\sqrt{19 \pi} \sigma^2)^{-1}$, $\beta=9(16\sqrt{19 \pi} \sigma^2)^{-1}$, and $q_o=-0.9\sqrt{3}\sigma$ where $\sigma=1.1\text{x}10^{-11}\text{m}$.}
	\label{leadingcorrection}
\end{figure*}

\begin{figure*}
	\includegraphics[width=\textwidth]{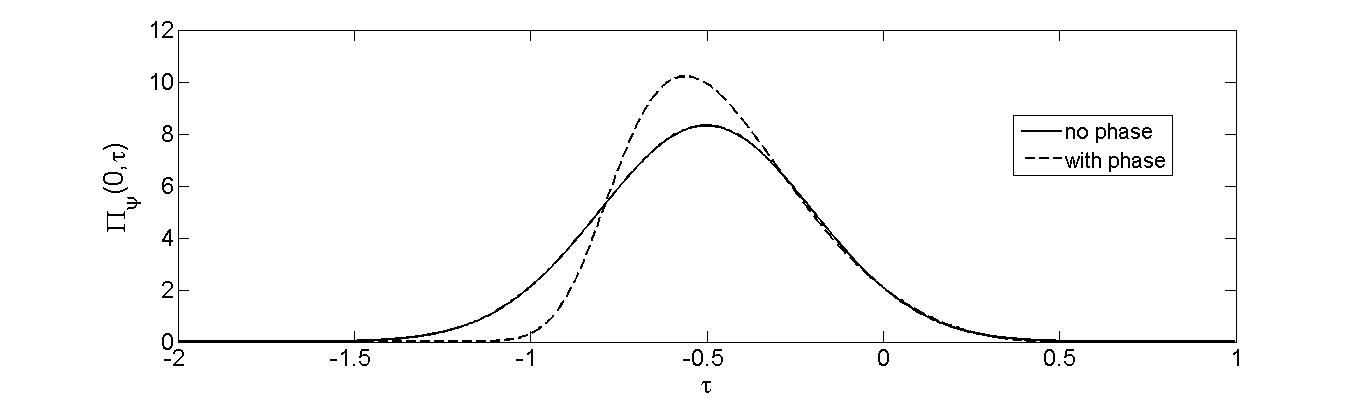}
	\caption{Quantum time of arrival distribution of neutrons for $\hbar=1$ with energy $E_0=200$, $\sigma=6$, $q_0=-10$, $\alpha=(32 \sqrt{6\pi})^{-1}$, and $\beta=5(576\sqrt{2\pi})^{-1}$.}
	\label{toadist}
\end{figure*}

Suppose we now want a phase that will make the second order quantum correction term vanish so that we get a leading quantum correction of the order $O(\hbar^3)$ to the classical time of arrival. This requires the condition $\chi_1^{(1)}=0$ or 
\begin{equation}
\label{cond3}
\int_{-\infty}^{+\infty}\;q\bigg(\frac{d\vartheta}{dq}\bigg)^{2}\Phi(q)dq+\int_{-\infty}^{+\infty}\;q\bigg(\frac{d}{dq}\sqrt{\Phi(q)}\bigg)^{2}dq = 0 .
\end{equation} 
We exploit the linearity of Eqs. \eqref{cond1} and \eqref{cond2} to simultaneously eliminate the $\mathcal{O}(\hbar)$ and $\mathcal{O}(\hbar^2)$ correction terms. Because of the linearity of \ref{cond1} and \ref{cond2}, any linear combination of Eqs. \eqref{odd} and \eqref{even} will remain to satisfy Eqs. \eqref{cond1} and \eqref{cond2}. Then we assume a phase of the form
\begin{equation}\label{phase2}
\vartheta(q)=\alpha \vartheta_{odd}(q)+\beta\vartheta_{even}(q)
\end{equation}
for some real constants $\alpha$ and $\beta$. Given one of the constants, the other can be determined by substituting Eq. \eqref{phase2} back into Eq. \eqref{cond3}, provided, of course, the solutions are real. In Section-\ref{example}, we will consider a specific example demonstrating the explicit construction of a phase that eliminates quantum corrections up to $\mathcal{O}(\hbar^2)$.

The method just described can be extended to eliminate an arbitrary oder of quantum corrections. However, the conditions beyond the first order correction are non-linear in phase so that an analytical construction of the required phase may be, in general, intractable to accomplish.

\section{Implementation of the quantum clock}\label{implementation}
To construct the quantum clock using quantum particles whose time interval marker is the QTOA of the particle, we need to address two issues. Namely, we first need to address the issue on a how the quantum particle can acquire a position dependent phase $\vartheta(q)$ , and secondly the fluctuations of the hand of the quantum clock around its mean position should be finite such that the width of the QTOA distribution is also finite. Here we consider the phase imprinting method and in the next Section we will consider a specific example addressing the second issue. 

To address the first, we can have the position dependent phase $\vartheta(q)$ imprinted onto the initial wave-function of  our quantum particle $\psi(q,t=0)=\psi_{np}(q)$ such that after the phase imprinting process, our final wave-function is given by $\psi(q,t=T)=\psi_{wp}(q)$. To do this, we use an impulsive interaction Hamiltonian to evolve the object. Its explicit form is given by
\begin{equation}
H(t)=- \gamma \Theta(q) \delta(t),
\label{potential}
\end{equation}
where $\gamma$ is a constant, $\Theta(q)$ is a real-valued function of $q$, and  $\delta(t)$ is a Dirac delta function in time. This form of the potential is chosen so that the phase imprinting process will only involve turning the potential on then immediately turning it off, with the phase of the form $\vartheta(q)=\gamma\Theta(q)/\hbar$ immediately afterwards.

\begin{figure*}
	\includegraphics[width=\textwidth]{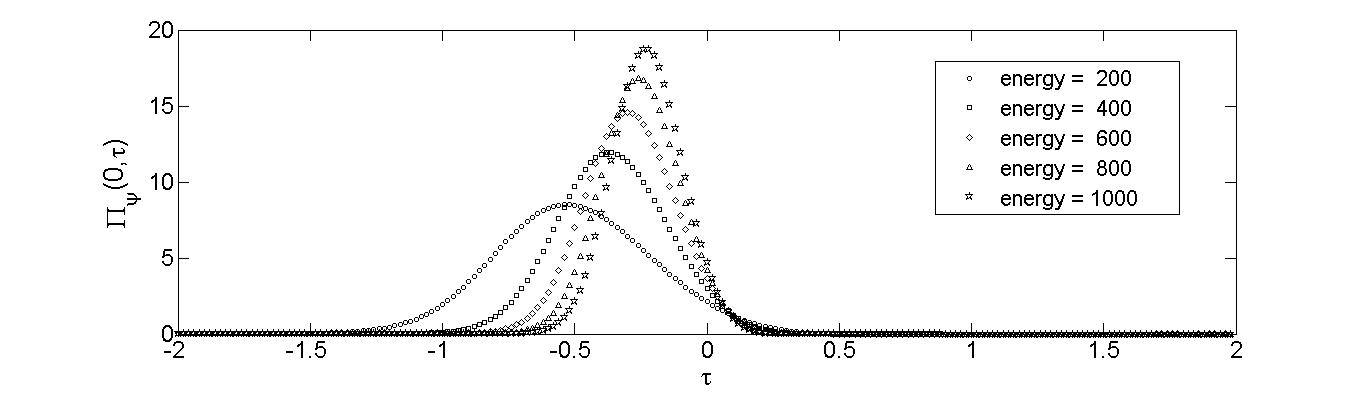}
	\caption{Quantum time of arrival distribution of neutrons with a position dependent phase and varying energies as indicated in the figure, for $\hbar=1$, $\sigma=6$, $q_0=-10$, $\alpha=(32 \sqrt{6\pi})^{-1}$, and $\beta=5(576\sqrt{2\pi})^{-1}$.}
	\label{toadistvarener}
\end{figure*}

Since the interaction Hamiltonian is time-dependent, the time evolution of the object's wave-function is described by the time-dependent Schr\"{o}dinger equation subject to the initial conditions stated above. The solution of this equation is given by
\begin{equation}
\ket{\psi(q,T)}=U(T,0)\ket{\psi(q,0)},
\label{finalstate}
\end{equation} 
where $\ket{\psi(q,0)}$ is the object's initial wave-function and $U(T,0)$ is the time evolution operator, which has the form
\begin{align}
U(T,0)=& \textrm{exp}\bigg(-\frac{i}{\hbar}\int_{0}^{T}H(t)dt\bigg) \nonumber \\
=& \textrm{exp}\bigg(\frac{i}{\hbar} \gamma \Theta(q)\bigg).
\label{timeevo}
\end{align}
From Eq. \eqref{timeevo} we see that after the phase imprinting process, the object's wave-function will have an additional phase $\vartheta(q)=\gamma \Theta(q) / \hbar$.

\section{Example}\label{example}
Now let us consider a specific example and demonstrate how the addition of phase can bring the expected QTOA closer to the classical time of flight. We use the particular values $l=0$ and $m=1$ in the phase given by Eq. \eqref{phase2}. Substituting the phase back into Eq. \eqref{cond3}, we obtain the equation
\begin{equation}
16 \pi  \sigma ^2 \left(4 \sqrt{3} \alpha  \beta  \sigma +q_0 \left(\alpha ^2+4 \beta ^2\right)\right)+\frac{q_0}{4 \sigma ^2}=0.
\label{coefficients}
\end{equation}
An inspection of Eq. \eqref{coefficients} suggests that $\alpha$ and $\beta$ has dimension $\text{length}^{-2}$. To be consistent with the dimensions we let $\alpha =a/\sigma^2$ and $\beta =b/\sigma^2$ where $a$ and $b$ are dimensionless quantities. Supposing we know $b$, $a$ is given by
\begin{equation}
\label{alpha}
a_{\pm}=\pm\frac{\sqrt{\pi } \sqrt{768 \pi  b ^2 \sigma ^2 - 256 \pi  b ^2 q_0^2 - q_0^2}-16 \sqrt{3} \pi  b  \sigma}{8 \pi  q_0}.
\end{equation}
However, we have imposed earlier that the position dependent phase $\vartheta(q)$ is a real-valued function. This means that $a$ and $b$ must both be real numbers and in order for this condition to be satisfied the parameter $b$ must satisfy the following inequality:
\begin{equation} 
\label{beta}
b^2\geq\dfrac{1}{1-\frac{1}{3}\frac{q_0^2}{\sigma^2}}\dfrac{q_o^2}{768\pi\sigma^2}
\end{equation}
where the parameters $\sigma$ and $q_0$ must satisfy:
\begin{equation}
\label{sigma}
\sigma^2>\dfrac{1}{3}q_0^2.
\end{equation}
By appropriately choosing the value of the parameters $b$, $\sigma$ and $q_0$ such that the inequalities in Eqs. (\ref{beta}) and (\ref{sigma}) are satisfied, an explicit form of the position dependent phase that will make the quantum correction terms vanish up to the second order and will make the leading quantum correction to the classical time of arrival in the order of $O(\hbar^3)$ exists. 


To better see the effect of imprinting a position dependent phase onto the initial wave-function of a quantum particle, we cast Eq. \eqref{asymptotic} into the following form
\begin{equation} 
\bar{\tau}_{quant} \sim   \tau_{class} \left(\sum_{n=0}^{\infty} \frac{\tilde{\chi}_1^{(n)}}{(k\sigma)^{2n}} -\frac{\sigma}{q_o}\sum_{n=0}^{\infty} \frac{(-1)^{n}\tilde{\chi}_2^{(n)}}{(k\sigma)^{2n+1}} \right)
\label{asymptotic2}
\end{equation}
where we have performed a change of variables from $q/\sigma=x+q_o/\sigma$ in Eq. \eqref{chi1} and Eq. \eqref{chi2} such that 
\begin{eqnarray}
\tilde{\chi}_1^{(n)}=\int_{-\infty}^{+\infty} |\tilde{\varphi}^{(n)}\!(x)|^2 dx,
\label{tildechi1}
\end{eqnarray}
\begin{eqnarray}
\tilde{\chi}_2^{(n)}&=&\int_{-\infty}^{+\infty} \left(x+\frac{q_o}{\sigma}\right) \, \mbox{Im}\!\left[\bar{\tilde{\varphi}}(x) \tilde{\varphi}^{(2n+1)}\!(x)\right] dx,
\label{tildechi2}
\end{eqnarray}
and
\begin{equation}
\tilde{\varphi}(x)=\frac{1}{(2 \pi)^{1/4}}e^{-x^2/4}e^{i \vartheta(x)}.
\end{equation}
We want to emphasize that by casting $\bar{\tau}_{quant}$ as Eq. \eqref{asymptotic2}, calculating the expectation value of the QTOA using  is now much more convenient than Eq. \eqref{asymptotic}. The integrands in Eq. \eqref{tildechi1} and \eqref{tildechi2} are now independent of $\sigma$ and $q_o$, making 
$\tilde{\chi}_1^{(n)}$ and $\tilde{\chi}_2^{(n)}$ dimensionless quantities. 

It is now easy to see that the leading quantum correction term for the classical time of arrival is equal to the classical TOA multiplied by a factor $Q(k \sigma)$. From Apendix II, we see that for $\psi_{np}$ the leading quantum correction term is proportional to 
\begin{equation}
Q_{np}(k\sigma)=\frac{\tilde{\chi}_{1_{np}}^{(1)}}{(k\sigma)^2} = \frac{1}{(k\sigma)^2} \int_{-\infty}^{+\infty}\bigg(\frac{d}{dx}\sqrt{\tilde{\Phi}(x)}\bigg)^{2}dx
\label{npleading}
\end{equation}  
while the leading quantum correction term for $\psi_{wp}$ is proportional to
\begin{equation}
Q_{wp}(k\sigma)=\frac{\sigma}{q_o}\frac{\tilde{\chi}_{2_{wp}}^{(1)}}{(k\sigma)^3}
\end{equation}
or equivalently,
\begin{align} \nonumber
Q_{wp}(k\sigma)=&\frac{1}{(k\sigma)^3}\frac{\sigma}{q_o}\int_{-\infty}^{\infty}\left(x+\frac{q_o}{\sigma}\right) \frac{d^{3}\vartheta}{dx^{3}}\tilde{\Phi}(x)dx\\
&-\frac{1}{(k\sigma)^3}\frac{\sigma}{q_o}\int_{-\infty}^{\infty}\left(x+\frac{q_o}{\sigma}\right) \bigg(\frac{d\vartheta}{dx}\bigg)^{3}\tilde{\Phi}(x)dx.
\label{wpleading}
\end{align}

An experiment by Badurek et al. \cite{experiment} modeled wavepackets as Gaussians with $\sigma=1.1\text{x}10^{-10}\text{m}$. Using this, we can set our parameters as $\alpha=5(4\sqrt{19 \pi} \sigma^2)^{-1}$, $\beta=9(16\sqrt{19 \pi} \sigma^2)^{-1}$, and $q_o=-0.9\sqrt{3}\sigma$ such that the inequalities in Eqs. (\ref{beta}) and (\ref{sigma}) are satisfied. Substituting these values to Eq. \eqref{phase2} ensures that the second order quantum correction term for $\psi_{wp}$ vanishes so that the leading quantum correction term is now $\tau_{class}Q_{wp}$. In our case, we will consider thermal neutrons with $E=0.025 \text{eV}$ and epithermal neutrons with energy range $0.025 \text{eV}$ to $0.4 \text{eV}$. This corresponds to $k=1.1\text{x}10^{12} \text{m}^{-1}$ and $k=4.4\text{x}10^{12} \text{m}^{-1}$ for  $E=0.025 \text{eV}$ and $E=0.4 \text{eV}$, respectively.

A comparison of the leading quantum correction factor, $Q(k \sigma)$, for wave-functions $\psi_{np}(q)$ and $\psi_{wp}(q)$, is shown Figure \ref{leadingcorrection}. The value $Q_{np}$ is positive which means that the QTOA is greater than the classical TOA while the value $Q_{wp}$ negative which means that the QTOA is less than the classical TOA. This means that in our case, the quantum clock runs slower than the classical clock and by imprinting the position dependent phase Eq. \eqref{phase2} we are able to make the quantum clock run faster but closer to the classical clock. Also, the value $Q_{np}$ and $Q_{wp}$ decreases and approaches zero as energy increases. This phenomenon is expected since the expectation value of the time of arrival operator approaches the classical value as $(\hbar^2/\mu E_o)\rightarrow 0$. However, it is important to note that $Q_{wp}$ approaches zero faster than $Q_{np}$ indicating that at higher energies, the effect of imprinting a position dependent phase is more noticeable. Thus, at higher energies imprinting a position dependent phase that eliminates the first and second order quantum correction terms is recommended since the magnitude of the leading quantum correction term for $\psi_{wp}$ is lower than the leading quantum correction for $\psi_{np}$. By doing so, we are able to synchronize quantum and classical clocks which uses the quantum and classical TOA as time interval markers, respectively, so that we get a more accurate calculation of the time of flight in TOF experiments.

We now demonstrate that the width of the QTOA distribution is finite so that the fluctuations of the hand of the quantum clock around its mean position is also finite. To do this, we use the nodal and non-nodal eigenfunctions of the standard time-of-arrival operator in momentum representation given by 
\begin{equation}
\Pi_{\psi_0}(X,\tau)=\abs{\braket{\psi_0}{\tau_{non}^X}}^2+\abs{\braket{\psi_0}{\tau_{nod}^X}}^2
\label{distribution}
\end{equation}
where
\begin{align}
\label{nonnodal}
\braket{p}{\tau_{non}^X}=&\sqrt{\frac{\abs{p}}{2\mu}}\frac{\exp^{-i p X / \hbar}}{\sqrt{2\pi\hbar}}\exp^{i p^2 \tau / 2 \mu \hbar} \\
\label{nodal}
\braket{p}{\tau_{nod}^X}=&\sqrt{\frac{\abs{p}}{2\mu}}\frac{\exp^{-i p X / \hbar}}{\sqrt{2\pi\hbar}}\exp^{i p^2 \tau / 2 \mu \hbar}\sign{p}.
\end{align}
The non-nodal contribution corresponds to particle appearance at the arrival point $X$ while the nodal contribution corresponds to a quantum arrival with no particle appearance \cite{denny2, giannitrapani}. 

The QTOA distribution at the arrival point $X=0$ for $\psi_{np}$ and $\psi_{wp}$ are plotted in Figure \ref{toadist} using Eqs. (\ref{distribution}) to (\ref{nodal}). Here we use a neutron and set $\hbar=1$. It can easily be seen in Figure \ref{toadist} that the width of the QTOA distribution is finite making the fluctuations of the hand of the quantum clock around its mean position to be also finite. We want to emphasize that this width is the full width at half maximum (FWHM) $\sigma_{\text{FWHM}}$ and not the square root of the variance of the QTOA distribution $\sigma_{\text{quant}}$,  
\begin{equation}
\sigma_{\text{quant}} = \sqrt{\bra{\phi}\opr{T}^2\ket{\phi}-\bra{\phi}\opr{T}\ket{\phi}^2} 
\label{sigmaquant}
\end{equation}
where $\opr{T}$ is the quantum arrival time operator for a free particle given in Eq. \eqref{freetoa}. The reason is that generally, $\sigma_{\text{quant}}$ is infinite because $\bra{\phi}\opr{T}^2\ket{\phi}$ does not always yields a finite value. A finite value for $\sigma_{\text{quant}}$ can only be achieved if the wave-function $\phi(q)$ is inside the domain of the operator $\opr{T}$ and in our case the wave-function $\phi(q)$ does not lie inside the domain of $\opr{T}$ \cite{bunao}.   

As such, a more apt parameter that we can use to characterize the spread of the distribution is its FWHM, which is finite. It is important to note that although the TOA distributions are skewed but using the FWHM it can be seen that in Figure \ref{toadist}, the width of the TOA distribution for $\psi_{wp}$ is smaller than that of $\psi_{np}$. This indicates that by imprinting a position dependent phase of the form given by Eq. \eqref{phase2} the width of the TOA distribution becomes smaller. 

Due to the finiteness of the FWHM, it is possible to set the parameters of the wave-function in such a way that the following condition is satisfied:
\begin{equation}
\sigma_{FWHM}/\bar{\tau}_{quant}<<1.
\label{antifluccond}
\end{equation}
If this condition is satisfied, then the fluctuations of the quantum clock's hand about its mean position can be minimized. One possible realization of this condition can occur in the high energy limit, where the fluctuations will be minimized
due to the narrowness of the arrival time distribution of the
particle. To illustrate how this can occur, first consider the plots of the arrival time probability distribution in Figure \ref{toadistvarener}. The plots show that for increasing energy, the distributions narrow in width, and consequently, FWHM will also decrease. 

\section{Summary}\label{summary}
We have shown that for any arbitrary quantum particle with velocity $v_o$ and whose initial wave-function is given by a quantum wave-packet moving through a potential-free region, its resulting quantum arrival time at a given point can be made equal to its corresponding classical arrival time $t=-q_o /v_o$ at $q_o$ up to a given order $N$ of $\hbar$. This can be done by imprinting onto the particle's initial wave-function a position-dependent phase $\vartheta(q)$ using an impulsive interaction Hamiltonian given by $H(t)=- \gamma \Theta(q) \delta(t)$. The form of the position-dependent phase $\vartheta(q)$ determines up to what order $N$ of $\hbar$ will the quantum correction terms in the particle's quantum arrival time vanish. Specifically, we have shown a form of the position dependent phase that can be imprinted so that the leading quantum correction term to the classical time of arrival is at most to the order of $O(\hbar^3)$. We acknowledge though that engineering the required phases is an entirely different problem which we leave open for future consideration.

\section{Acknowledgments}

The authors would like to acknowledge A. Villanueva for useful comments and suggestions in the initial stages of drafting this paper, and D.L. Sombillo during the final stages. 

\section*{Appendix I: Derivation of the Full Asymptotic Expansion of the TOA Expectation Value}

In this Appendix we recap, for completeness, the derivation of Eq. \eqref{asymptotic}, the full asymptotic form of the TOA expectation value as $\hbar\rightarrow 0$, which was carried out in Ref. \cite{gal2}. Our starting point is the quantum expected time of arrival for the wave-function $\phi(q)$ given by
\begin{eqnarray}
\bar{\tau}_{quant}\!\!&=&\!\!\int_{-\infty}^{\infty}\int_{-\infty}^{\infty}\!\left[ \frac{\mu}{4 i \hbar}(q+q')\, \mbox{sgn}(q-q')\right]\nonumber \\
&& \hspace{12mm}\times \bar{\varphi}(q)
\varphi(q') e^{-i k(q- q')} dq' dq. 
\end{eqnarray}
This expression can be cast into the form 
\begin{eqnarray}
\bar{\tau}_{quant}=\int_{-\infty}^{\infty}\bar{\phi}(q)  \psi_{k}(q) \, dq,\label{bebe}
\end{eqnarray}
where
\begin{eqnarray}
\psi_k(q)\!\!&=&\!\!\frac{\mu}{4 i\hbar}\int_{-\infty}^{\infty} (q+q')\, \mbox{sgn}(q-q')\varphi(q') e^{i k q'}\, dq' \nonumber \\
&=& \frac{\mu}{4 i\hbar}\int_{-\infty}^{q} (q+q')\, \varphi(q') e^{i k q'}\, dq' \nonumber \\
& & \hspace{16mm}- \frac{\mu}{4 i\hbar}\int_{q}^{\infty} (q+q') \varphi(q') e^{i k q'}\, dq', \label{qu} \nonumber \\
\end{eqnarray} 
where we have used the identity $\text{sgn}(q-q')=H(q-q')-H(q'-q)$, in which $H(x)$ is the Heaviside function, to arrive at Eq.\ref{qu}.

Repeated integration by parts yields the full asymptotic expansion of the quantum expected time of arrival
\begin{eqnarray}
\psi_k(q)&\sim& \frac{\mu}{2 i \hbar} e^{i k q} \sum_{n=0}^{\infty} \frac{(-1)^n}{(ik)^{n+1}} \left.\frac{\partial^n}{\partial q'^n}\left[(q+q') \varphi(q')\right]\right|_{q'=q} \nonumber 
\end{eqnarray}
as $k\rightarrow\infty$. This expression holds for all $\varphi(q)$ that is rapidly decreasing, such as gaussian wavepackets. Evaluating the first few partial derivatives in the above expression, we find the pattern
\begin{eqnarray}
\frac{\partial^n}{\partial q'^n}\left[(q+q') \varphi(q')\right]=n \varphi^{(n-1)}(q') + (q+q') \varphi^{(n)}(q').\label{kaw}
\end{eqnarray}
We can prove that this expression is correct by induction, which is done by taking the partial derivative of the right hand side of Eq. \eqref{kaw} and then showing that it is equal with Eq. \eqref{kaw} with $n$ shifted to $(n+1)$. Then the asymptotic value of the quantum expected time of arrival is obtained by substituting the asymptotic expansion of $\psi_k(q)$ back into Eq. \eqref{bebe} and is given by
\begin{eqnarray}
\bar{\tau}_{\varphi}
\!\!& \sim &\!\! \frac{\mu}{2 i \hbar}\! \sum_{n=0}^{\infty}\!\!\frac{(-1)^n}{(ik)^{n+1}}\!\!\!\!\int_{-\infty}^{\infty}\!\!\!\!\!\!\!\!\bar{\varphi}(q)\!  (n\varphi^{(n-1)}\!(q)\! +\! 2 q \varphi^{(n)}\!(q)) dq \nonumber \\
\end{eqnarray}
as $k\rightarrow\infty$. An inspection of the first few terms of this series shows that even and odd powers of $k$ must be treated separately. 

For $n=2j$, $j=1,2,\dots$, we consider the integral $I_{1 }^{(j)}=\int_{-\infty}^{\infty} \bar{\varphi}(q) q \varphi^{(2j)}(q)\, dq$ and perform the following manipulations,
\begin{eqnarray}
I_1^{(j)}
\!\!\!&=&\!\!\! \int_{-\infty}^{\infty}\!\!\! \bar{\varphi}(q) \frac{d}{dq}\left(q \varphi^{(2j-1)}(q)\right)-\int_{-\infty}^{\infty}\!\!\! \bar{\varphi}(q) \varphi^{(2j-1)}(q)\, dq \nonumber \\
\!\!\!&=&\!\!\! -\int_{-\infty}^{\infty}\!\!\!\!\! \bar{\varphi}^{(1)}(q)q \varphi^{(2j-1)}(q)\,dq-\int_{-\infty}^{\infty}\!\!\!\! \bar{\varphi}(q) \varphi^{(2j-1)}(q)\, dq \nonumber .
\end{eqnarray}
The first term of the last line follows from the first term of the second term after performing an integration by parts. Observe that the first term of the last line has the same structure as $I_1^{(j)}$. Performing the same manipulation on this term, and continuing the process $j$-times leads to the identity
\begin{eqnarray}
\int_{-\infty}^{\infty}q\, \bar{\varphi}\!(q) \varphi^{(2j)}\!(q) dq &=& (-1)^j \int_{-\infty}^{\infty} q \abs{\varphi^{(j)}\!(q)}^2 dq \nonumber \\
& & \hspace{-12mm} - j \int_{-\infty}^{\infty} \bar{\varphi}\!(q) \varphi^{(2j-1)}\!(q) dq.  \nonumber
\end{eqnarray}
For $n=(2j+1)$ let us consider the integral $I_2^{(j)}=\int_{-\infty}^{\infty}\bar{\varphi}(q) q \varphi^{(2j+1)}(q) dq$. Perfoming similar manipulations as above for $(2j+1)$-times yields the integral identity
\begin{eqnarray}
\int_{-\infty}^{\infty}\!\!\!\!\! q \,\!\bar{\varphi}(q) \varphi^{(2j+1)}  dq &=&-\int_{-\infty}^{\infty}\!\!\!\!\! q \,\!\varphi(q) \bar{\varphi}^{(2j+1)}  dq\nonumber \\
&& \hspace{-12mm} - (2j+1)\!\!\!\int_{-\infty}^{\infty}\!\!\! \bar{\varphi}\!(q) \varphi^{(2j)}(q) dq. \nonumber
\end{eqnarray}

Substituting these two integrals back yields the full asymptotic form of the quantum expected time of arrival
\begin{eqnarray}
\bar{\tau}_{quant}\!\! &\sim &\!\!  - \frac{\mu}{\hbar}\sum_{j=0}^{\infty} \frac{\chi_1^{(j)}}{k^{2j+1}}   + \frac{\mu}{\hbar}\sum_{j=0}^{\infty}  \frac{(-1)^j \chi_2^{(j)}}{k^{2j+2}}\label{com} 
\end{eqnarray}
as $k\rightarrow\infty$. In terms of the kinetic energy $E_o$, we have
\begin{eqnarray}
\bar{\tau}_{quant}
\sim  - \sqrt{\frac{\mu}{2 E_o}}\!\! \sum_{n=0}^{\infty} \!\!\frac{\hbar^{2n} \chi_1^{(n)}}{(2\mu E_o)^n} 
\!+\! \frac{1}{2 E_o}\!\! \sum_{n=0}^{\infty}  \!\! \frac{(-1)^n\hbar^{2n+1} \chi_2^{(n)}}{(2\mu E_o)^{n}}. \label{asymp2}
\end{eqnarray}

\section*{Appendix II: Calculation of the Quantum Correction Terms}

In this Appendix, we calculate the quantum correction terms up to the sixth order and show that there is no general form for the $n^{th}$ order quantum correction term by showing that no observable pattern can be found for the quantum correction terms. We do this by substituting Eq. \eqref{wave1} to Eqs. \ref{chi1} and \ref{chi2}. The resulting quantum correction terms will then have the following form:
\begin{align}
\label{1storder}
\chi_{2}^{(0)}=\int_{-\infty}^{+\infty}\;q\frac{d\vartheta}{dq}\Phi(q)dq.
\end{align}
\begin{align}
\label{2ndorder}
\chi_{1}^{(1)}=\int_{-\infty}^{+\infty}\;q\bigg(\frac{d\vartheta}{dq}\bigg)^{2}\Phi(q)dq+\int_{-\infty}^{+\infty}\;q\bigg(\frac{d}{dq}\sqrt{\Phi(q)}\bigg)^{2}dq
\end{align}
\begin{align} 
\label{3rdorder}
&\chi_{2}^{(1)}&=&-\int_{-\infty}^{+\infty}\;q\bigg(\frac{d\vartheta}{dq}\bigg)^{3}\Phi(q)dq+\int_{-\infty}^{+\infty}\;q\frac{d^{3}\vartheta}{dq^{3}}\Phi(q)dq
\end{align}
\begin{align}
\label{4thorder} \nonumber
&\chi_{1}^{(2)}&=&\int_{-\infty}^{+\infty}\;q\bigg(\frac{d^{2}}{dq^{2}}\sqrt{\Phi(q)}\bigg)^{2}dq+\int_{-\infty}^{+\infty}\;q\bigg(\frac{d^{2}\vartheta}{dq^{2}}\bigg)^{2}\Phi(q)dq\\ 
&              & &+\int_{-\infty}^{+\infty}\;q\bigg(\frac{d\vartheta}{dq}\bigg)^{4}\Phi(q)dq. \\ \nonumber
\end{align}
\begin{align}
\label{5thorder} \nonumber
&\chi_{2}^{(2)}&=&\int_{-\infty}^{+\infty}\;q\frac{d^{5}\vartheta}{dq^{5}}\Phi(q)dq+\int_{-\infty}^{+\infty}\;q\bigg(\frac{d\vartheta}{dq}\bigg)^{5}\Phi(q)dq\\ \nonumber
&              & &-10\int_{-\infty}^{+\infty}\;q\bigg(\frac{d^{3}\vartheta}{dq^{3}}\bigg)\bigg(\frac{d\vartheta}{dq}\bigg)^{2}\Phi(q)dq\\ 
&              & &-15\int_{-\infty}^{+\infty}\;q\bigg(\frac{d\vartheta}{dq}\bigg)\bigg(\frac{d^{2}\vartheta}{dq^{2}}\bigg)^{2}\Phi(q)dq.
\end{align} 
\begin{align}
\label{6thorder} \nonumber
&\chi_{1}^{(3)}&=&\int_{-\infty}^{+\infty}\;q\bigg(\frac{d^{3}}{dq^{3}}\sqrt{\Phi(q)}\bigg)^{2}dq \\ \nonumber
&              & &-6\int_{-\infty}^{+\infty}\;q\sqrt{\Phi(q)}\frac{d^2 \vartheta(q)}{dq^2}\frac{d \vartheta(q)}{dq}\frac{d^{3}}{dq^{3}}\sqrt{\Phi(q)}dq \\ \nonumber
&              & &+9\int_{-\infty}^{+\infty}\;q\bigg(\frac{d^2 \vartheta(q)}{dq^2}\bigg)^2\bigg(\frac{d \vartheta(q)}{dq}\bigg)^2 \Phi(q)dq \\ \nonumber
&              & &+\int_{-\infty}^{+\infty}\;q\bigg(\frac{d^3 \vartheta(q)}{dq^3}\bigg)^2\Phi(q)dq \\ \nonumber
&              & &+\int_{-\infty}^{+\infty}\;q\bigg(\frac{d\vartheta(q)}{dq}\bigg)^6\Phi(q)dq. \\ 
&              & &+2\int\limits_{-\infty}^{+\infty}q\frac{d^3\vartheta(q)}{dq^3}\bigg(\frac{d\vartheta(q)}{dq}\bigg)^3\Phi(q)dq. \\ \nonumber
\end{align}

As can be seen in Eqs. \ref{1storder} to \ref{6thorder}, the quantum correction terms become more complicated as its order increases and that there is no general form for the $n^{th}$ order correction term. This means that we can't find a general form for $\vartheta(q)$ such that all the quantum correction terms will vanish. However, it is important to note that the $n^{th}$ order correction term always has a term with $(\vartheta'(q))^n$ in its integral. If we assume a solution of the form $\vartheta(q)=\sum_{k=1}^{n+1}c_{k}\vartheta_{k}(q)$ this yields a nonlinear system of $n+1$ equations with $n+1$ unknowns and the $k^{th}$ order correction term becomes a polynomial of order $k$. Using this system of equations, the coefficients $c_k$ can be solved to make the correction terms vanish up to a certain order depending on the choice of $\vartheta(q)$ and at the same time satisfy Eq. \eqref{cond1}. 

\section*{Appendix III: Derivation of $\vartheta_{odd}(q)$ and $\vartheta_{even}(q)$}

In this Appendix we recap, for completeness, the derivation of Eqs. \ref{odd} and \ref{even} which was carried out in Ref. \cite{bunao}. Our starting point is the linear system of equations given by
\begin{equation} 
	\int_{-\infty}^{+\infty}\varphi(q)dq= 0 ,\int_{-\infty}^{+\infty}q \varphi(q)dq= 0.
	\label{linearsys}
\end{equation}
Assume $\varphi(q)=\alpha\varphi_1(q)+\beta\varphi_2(q)$ where  $\varphi_1(q)$ and $\varphi_2(q)$ are linearly independent functions, and $\alpha$ and $\beta$ are constants such that conditions \ref{linearsys} are satisfied. Substituting  $\varphi(q)$ to Eq. \eqref{linearsys} we obtain the following matrix expression for $\alpha$ and $\beta$,
\begin{equation}
	\begin{bmatrix}
		\int_{-\infty}^{+\infty}\varphi_1(q)dq & \int_{-\infty}^{+\infty}\varphi_2(q)dq \\
		\int_{-\infty}^{+\infty}q\varphi_1(q)dq & \int_{-\infty}^{+\infty}q\varphi_2(q)dq
	\end{bmatrix}
	\begin{bmatrix}
		\alpha \\
		\beta
	\end{bmatrix}
	=0.
\end{equation} 

A unique solution for $\alpha$ and $\beta$ exists if the determinant of the matrix of the coefficients vanishes which gives the following condition for the integrals of $\varphi_1(q)$ and $\varphi_2(q)$,
\begin{align} \nonumber
	&\left(\int_{-\infty}^{+\infty}\varphi_1(q) dq\right)\left(\int_{-\infty}^{+\infty}q\varphi_2(q)dq\right) \\
	&=\left(\int_{-\infty}^{+\infty}\varphi_2(q)dq\right)\left(\int_{-\infty}^{+\infty}q\varphi_1(q)dq\right).
	\label{determinant}
\end{align}
Eq. \eqref{determinant} can be satisfied by letting $\varphi_1(q)$ and $\varphi_2(q)$ have definite parities, i.e. either both of them are odd or even. With these conditions, we can write:
\begin{align} \nonumber  \varphi_{o}(q)=&\left(\int_{-\infty}^{+\infty}y\varphi_1(y) dy\right)\varphi_2(q) \\
	&-\left(\int_{-\infty}^{+\infty}y\varphi_2(y) dy\right)\varphi_1(q).
	\label{varphiodd}
\end{align}
for odd $\varphi_1(q)$ and $\varphi_2(q)$, and 
\begin{align} \nonumber
	\varphi_{e}(q)=&\left(\int_{-\infty}^{+\infty}\varphi_1(y) dy\right)\varphi_2(q)\\
	&-\left(\int_{-\infty}^{+\infty}\varphi_2(y) dq\right)\varphi_1(q)
	\label{varphieven}
\end{align}
for even $\varphi_1(q)$ and $\varphi_2(q)$. 

If we now choose $\varphi_k(q)=(2^{k/2}\sqrt{k!}\sqrt[4]{\pi})^{-1}\;H_k(q)e^{-q^2/2}$,  where $k=0,1,\dots$ and $H_k(q)$'s are the Hermite polynomials. Each $\varphi_k(q)$ has a definite parity depending on $k$, i.e. $\varphi_k(q) $ is odd (even) for odd (even) $k$.    We can now define $\varphi(q)$ satisfying Eq. \eqref{linearsys} as 
\begin{align} \nonumber
	\varphi_{o}(q)=&\dfrac{\sqrt{2}}{2^l 2^m l!}\sqrt{\dfrac{(2l+1)!}{(2m+1)!}}H_{2m+1}(q)e^{-q^2/2}\\
	&-\dfrac{\sqrt{2}}{2^l 2^m m!}\sqrt{\dfrac{(2m+1)!}{(2l+1)!}}H_{2l+1}(q)e^{-q^2/2}
\end{align} 
and
\begin{align} \nonumber
	\varphi_{e}(q)=&\dfrac{\sqrt{2}}{2^l 2^m l!}\sqrt{\dfrac{(2l)!}{(2m)!}}H_{2m}(q)e^{-q^2/2}\\
	&-\dfrac{\sqrt{2}}{2^l 2^m m!}\sqrt{\dfrac{(2m)!}{(2l)!}}H_{2l}(q)e^{-q^2/2}
\end{align} 
where $l\neq m$.

By comparing Eqs. (\ref{linear}) and (\ref{linearsys}) we see that $\varphi(q)=\Phi(q)\vartheta'(q)$ where $\Phi(q)$ is given by Eq. \eqref{density}. By making a change of variables in Eqs. (\ref{varphiodd}) and (\ref{varphieven}) from $q$ to $(q-q_0)/\sigma$ it is easy to see that
\begin{align} \nonumber
	\dfrac{d\vartheta_{even}(q)}{dq}=&\dfrac{2\sigma\sqrt{\pi}}{2^l 2^m l!}\sqrt{\dfrac{(2l+1)!}{(2m+1)!}}H_{2m+1}(\frac{q-q_{0}}{\sigma})\\
	&-\dfrac{2\sigma\sqrt{\pi}}{2^l 2^m m!}\sqrt{\dfrac{(2m+1)!}{(2l+1)!}}H_{2l+1}(\frac{q-q_{0}}{\sigma})
	\label{deven}
\end{align}
\begin{align} \nonumber
	\dfrac{d\vartheta_{odd}(q)}{dq}=&\dfrac{2\sigma\sqrt{\pi}}{2^l 2^m l!}\sqrt{\dfrac{(2l)!}{(2m)!}}H_{2m}(\frac{q-q_{0}}{\sigma})\\
	&-\dfrac{2\sigma\sqrt{\pi}}{2^l 2^m m!}\sqrt{\dfrac{(2m+1)!}{(2l+1)!}}H_{2l}(\frac{q-q_{0}}{\sigma}).
	\label{dodd}
\end{align}
By integrating Eqs. (\ref{deven}) and $\ref{dodd}$ we get Eqs. (\ref{even}) and (\ref{odd}), respectively.


\end{document}